\documentclass[journal,comsoc]{IEEEtran}

\usepackage[T1]{fontenc}

\usepackage{cite}
\usepackage{color}
\usepackage{soul}

\ifCLASSINFOpdf
  \usepackage[font=footnotesize, labelfont={sf,bf}, margin=1cm]{caption}
  \usepackage[pdftex]{graphicx}
  \graphicspath{{../pdf/}{../jpeg/}}
  \DeclareGraphicsExtensions{.pdf,.jpeg,.png}
\else

\fi

\usepackage{amsmath}
\usepackage{gensymb}
\usepackage{subfigure}
\usepackage{balance}

\let\svthefootnote\thefootnote

\begin{document}

\title{A Software-Defined Multi-Element VLC Architecture}


\author{\IEEEauthorblockN{Sifat Ibne Mushfique, Prabath Palathingal, Yusuf Said Ero\u{g}lu, Murat Y\"{u}ksel}, \textit{Senior Member, IEEE,} \\  {\.I}smail G\"{u}ven\c{c}, \textit{Senior Member, IEEE} and Nezih~Pala}
	
%
%
%
	

\maketitle


\begin{abstract}
In the modern era of radio frequency (RF) spectrum crunch, visible light communication (VLC) is a recent and promising alternative technology that operates at the visible light spectrum. Thanks to its unlicensed and large bandwidth, VLC can deliver high throughput, better energy efficiency, and low cost data communications. In this article, a hybrid RF/VLC architecture is considered that can simultaneously provide lighting and communication coverage across a room. Considered architecture involves a novel multi-element hemispherical bulb design, which can transmit multiple data streams over light emitting diode (LED) modules. Simulations considering various VLC transmitter configurations and topologies show that good link quality and high spatial reuse can be maintained in typical indoor communication scenarios.
\end{abstract}


\begin{IEEEkeywords}
Free space optics (FSO), light emitting diode (LED), visible light communication (VLC).
\end{IEEEkeywords}

\section{Introduction}
%
%
%
%
\let\thefootnote\relax\footnote{Sifat Ibne Mushfique and Murat Y\"{u}ksel are with the Department of Electrical and Computer Engineering at University of Central Florida, Orlando, FL. \{email: sifat.im@knights.ucf.edu, murat.yuksel@ucf.edu\}}
\footnote{Prabath Palathingal is with Raven Electronics Corporation,  4655 Longley Lane, Reno, NV. \{email: prabath.palathingal@ravencomm.com\}}
\footnote{Yusuf Said Ero\u{g}lu and {\.I}smail G\"{u}ven\c{c} are with the Department of Electrical and Computer Engineering at the North Carolina State University, Raleigh, NC. \{email: yeroglu@ncsu.edu, iguvenc@ncsu.edu\}}
\footnote{Nezih Pala is with the Department of Electrical and Computer Engineering at Florida International University, Miami, FL. \{email: npala@fiu.edu\}}
\footnote{This work is supported in part by NSF CNS awards 1422354 and 1422062.}
\addtocounter{footnote}{-5}\let\thefootnote\svthefootnote
Multi-element visible light communication (VLC) has been recently receiving extensive interest as a new paradigm that can simultaneously maintain desirable communication properties such as high speed and long range, as well as 
high and even intensity of illumination. 
The directional beams in VLC systems, while requiring line of sight (LOS) connectivity, open up great opportunities for spatial reuse of optical spectrum resources. Multi-element VLC modules can significantly improve the efficiency of data transmission as they can take full advantage of directional property of light by modulating each transmitter (e.g., a light emitting diode (LED)) with a different data stream. By
designing these multi-element modules conformal to spherical shapes, one may also provide uniform light coverage
across the room. 

\begin{figure*}[!htbp]
	\centering
	\subfigure[Inefficient use of spatial resources in a single-element architecture.]{\includegraphics[width=80mm]{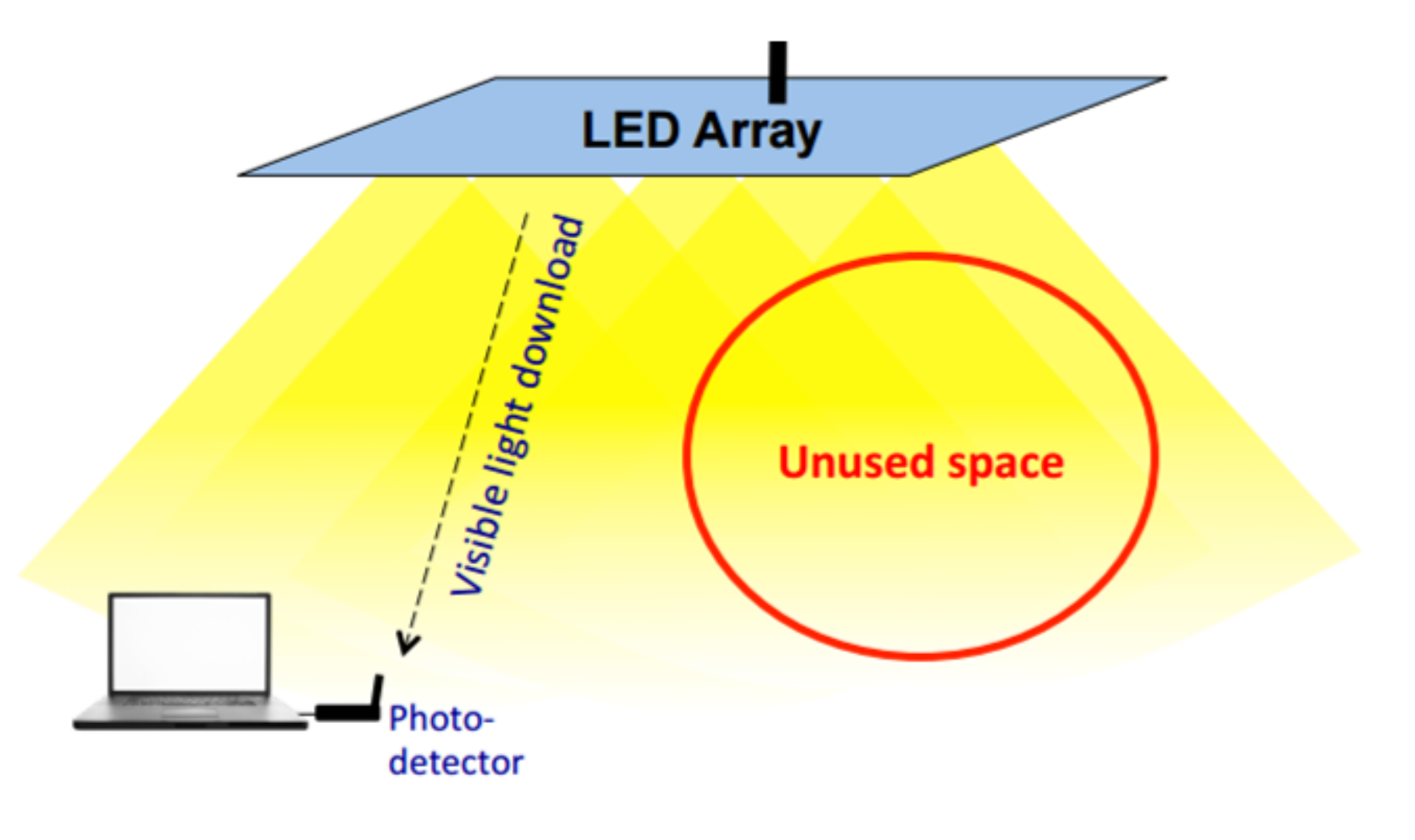}
		\label{fig_room}}
	\subfigure[Sample design of a multi-element VLC architecture.]{\includegraphics[width=80mm]{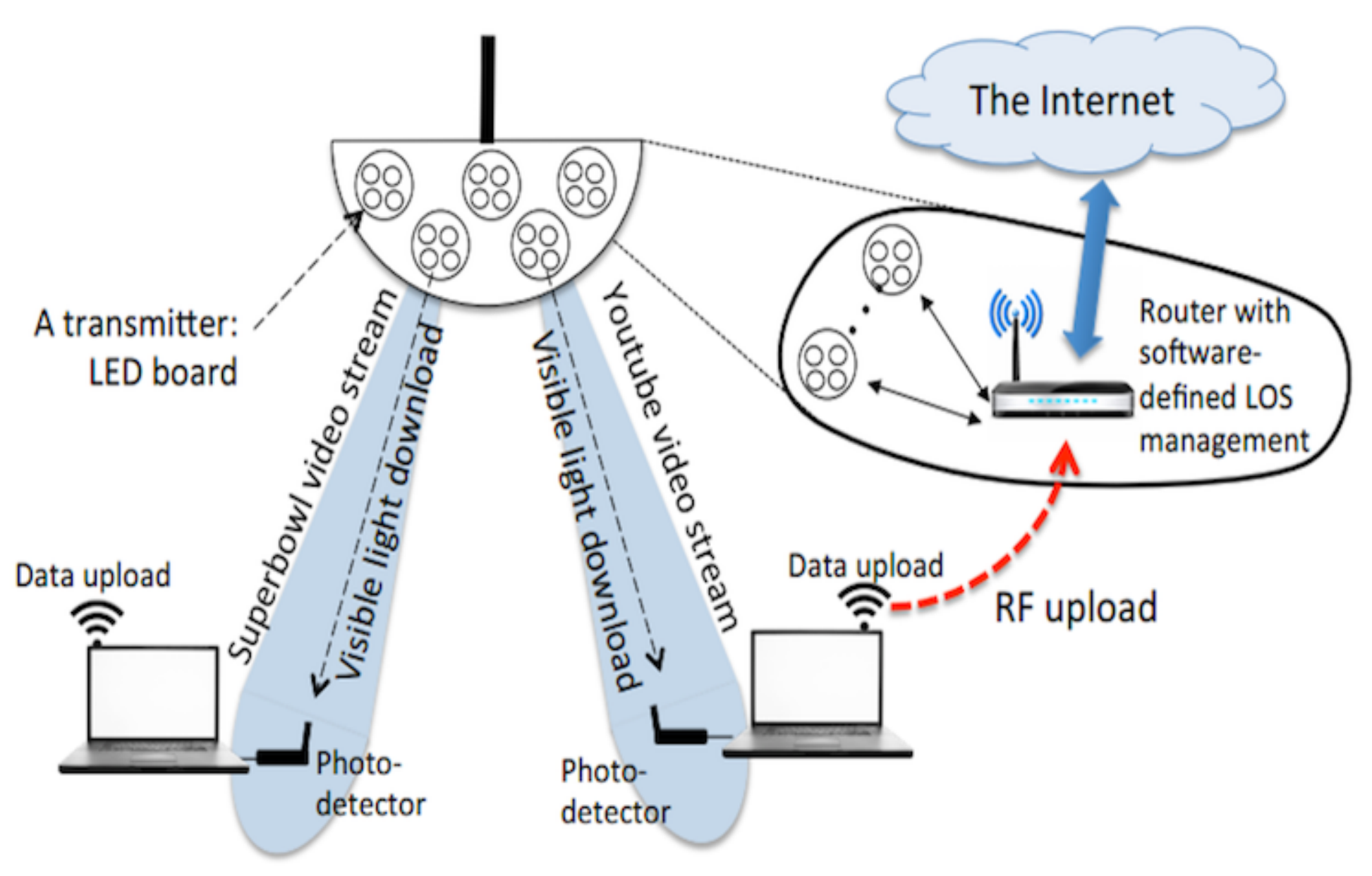}
		\label{fig_arch}}%
	\caption{Benefit of a multi-element VLC architecture.}
	\label{fig:multi-VLC}
\end{figure*}

The existing work in the literature related to multi-element VLC are on expanding the field-of-view (FOV), range, and rate of communication, and significant advances have been attained in increasing what is possible with a single element, i.e., a transmitter (an LED), receiver (a photo-detector (PD)), or transceiver
(an LED-PD pair) \cite{2001-willebrand-free, 2013-sevincer-lightnets}.
In~\cite{chen2014improving,Combining_Receiver_2014}, multi-element receiver approaches are introduced to improve system performance. There have been earlier works that jointly study illumination and communication aspects of VLC systems ~\cite{choi2014visible}, optical beamforming \cite{morrison2016directional} where devices such as micromirrors are used to dynamically improve an optical wireless communications link, and hybrid RF/FSO networks~\cite{shaoindoor} that  consider VLC in the down-link and RF communication (such as WiFi) in the up-link. Also, in~\cite{phpathak-survey, dilukshan-survey}, the main concept of the VLC architecture is presented in a way where light emitted from a single LED is focused in a specific target direction using spatial light modulator (SLM)~\cite{kim2013performance}.
However, in a single-element architecture, it is not possible to take advantage of LED directionality to achieve spatial reuse as shown in Figure \ref{fig_room}, which can be achieved using multi-element transceiver architectures.

Considering the limited functionality and efficiency of the single element VLC architecture, the major focus of this work is to explore designs using many elements with narrow FOV. In particular, we use multi-element VLC modules for simultaneous transfer of multiple data streams and attain higher spatial reuse in short ranges, (e.g., a room), due to the dense grid formed by the narrow FOV LEDs. Unlike the recent works, our research does not focus in the integration of multiple spotlighting mechanism into a single light source (base station), thereby creating an apparently large FOV which makes handover easier and also serves the purpose of illumination. Instead of developing a VLC-based broadcast system~\cite{paper-1-Dobrien}, our research focuses on unicast data stream from individual spotlight from an overhead light source. This approach will likely be more practical for the emerging Internet-of-Things (IoT) applications that involve closely placed receivers accessing the VLC resources. Further, since most of these IoT receivers will not be highly mobile, handover across spotlights will not be prohibitively costly.

We consider our proposed architecture to be an example of software-defined VLC systems that heavily employ software components (as
opposed to analog or mechanical components, e.g. for detection at the receiver and beam steering at
transmitter) to solve the inherent VLC problems such as alignment, LOS, and seamless
integration to legacy RF-based technologies. The proposed multi-element VLC architecture specifically uses software to handle the LOS
discovery and the alignment handling problems that are crucial to a mobile setting. These problems become
particularly crucial when LEDs have narrower divergence angles. The  architecture uses software protocols
to detect, maintain and optimize the LOS alignment to receivers in the room while maximizing spatial reuse
(for more aggregate throughput) as well as minimizing the variation in lighting quality. Further, we use
software to actively associate receivers to LED groups on the multi-LED bulb. Since the throughput requirement in the downlink is much higher than the uplink in typical indoor applications, we focus on the downlink VLC design as a complement to existing wireless technologies such as WiFi.

\section{Multi-Element Architecture}
\label{sec:multielement-architecture}

There are two main objectives of our multi-element
VLC approach: 1) \emph{high spatial reuse} by fully utilizing the
directionality of LEDs, and 2) \emph{seamless handling of mobility} of
receivers by using software protocols that steer the data
transmissions to mobiles. 
Unlike the traditional design of LEDs/transmitters with large divergence angles, we propose to use narrow divergence angles and still perform an acceptable illumination by using a large number of LEDs on a ``bulb''. In particular, our work - 1) advocates an architecture with LEDs with narrow divergence angles to reap the benefits of spatial reuse; 2) uses software protocols for seamless handling of mobility issues such as LOS discovery, alignment maintenance, and receiver association; and 3) utilizes software-based heuristic optimizations to solve interference problems between simultaneous VLC links, which is significantly different from the concepts described  in~\cite{phpathak-survey, dilukshan-survey}.

\begin{figure*}[!htbp]
	\centering
	\subfigure[Placement of transmitters.]{\includegraphics[width=2.4in, height=3.5in, keepaspectratio]{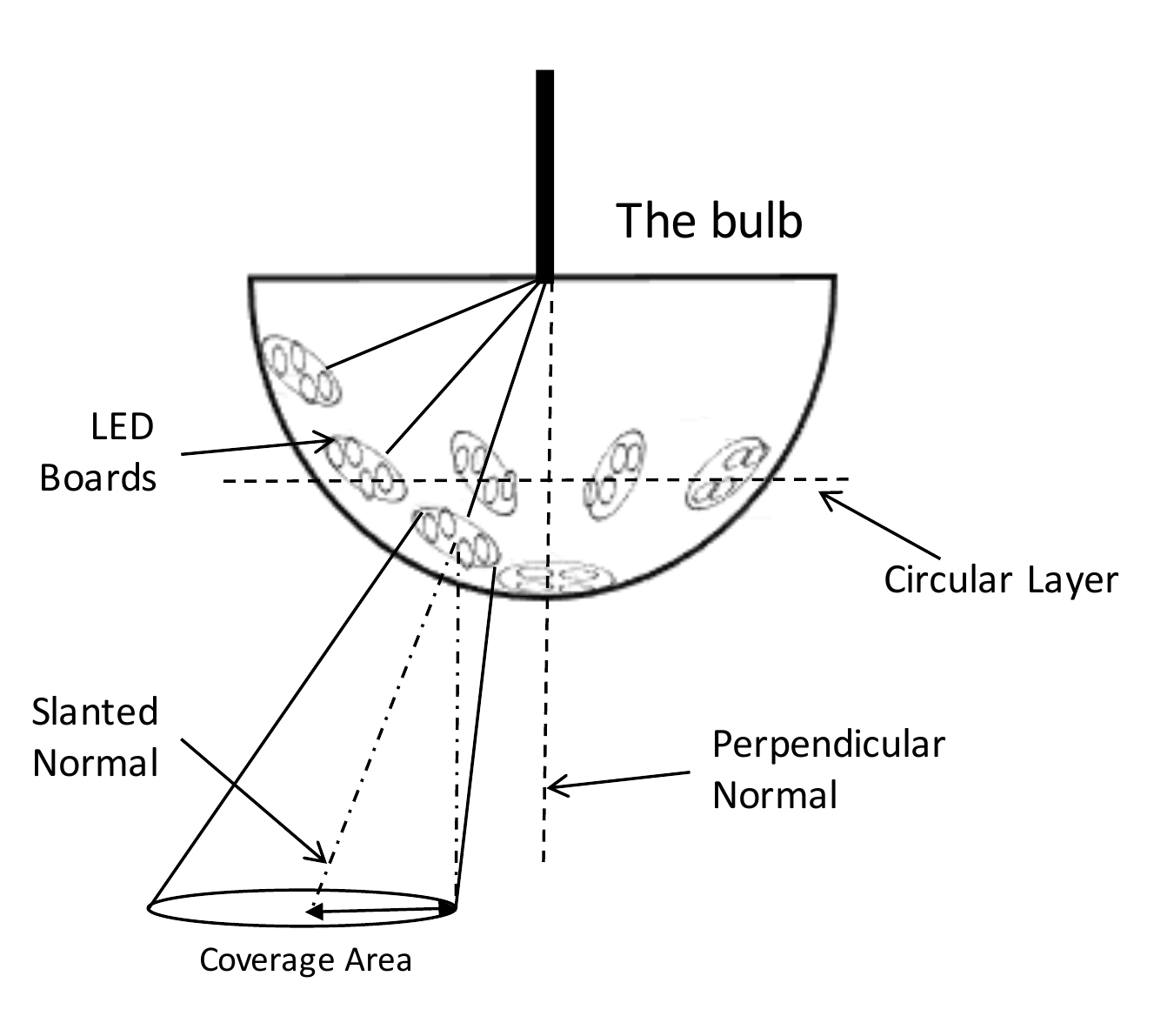}%
		\label{fig_tx_structure}}
	\hfil
	\subfigure[Partitioning algorithm.]{\includegraphics[width=3.2in, height=3.5in, keepaspectratio]{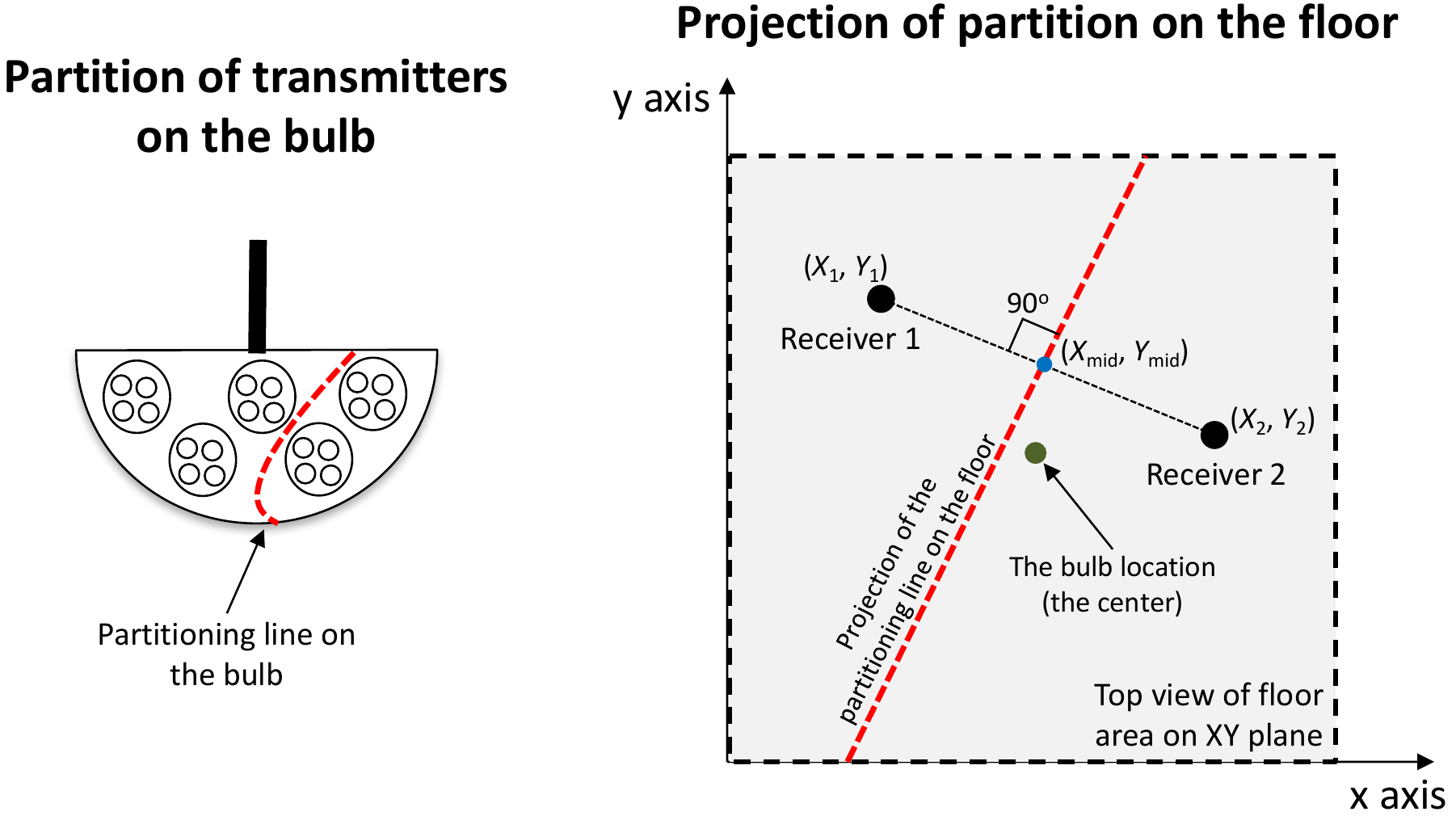}%
		\label{fig_part_alg}}
	\caption{Transmitters in the multi-element bulb.}
	\label{fig_tx_in_multi}
\end{figure*}

Even when we use hundreds of LEDs, we still have the problem of steering the data transmission to the corresponding LED when the mobile receiver is moving. We tackle this problem with a software-defined and enhanced version of electronic steering~\cite{2013-sevincer-lightnets}. Our architecture takes advantage of spatial reuse and seamless steering which are untapped sources of efficiency in VLC. We detail the architecture in Figure \ref{fig_arch} by describing three key components below.

\subsection{The Bulb}
The bulb is a hemispherical structure, which will act as an access point for the room. There are multiple transmitters in the bulb to provide \emph{concurrent downloads} to multiple receivers. A transmitter is basically an LED board which has a chunk of LEDs for transmitting a particular data stream. The LEDs on the same transmitter are all modulated by the same signal. For that reason, having multiple LEDs on a transmitter board allows operating in a wide range of configurations involving source power, communication range, and illumination quality. 
However, there remains the challenge of seamless steering of data to the corresponding transmitters. To address this issue, we connect each transmitter to a controller device 
embedded in the bulb and run a software protocol for managing LOS alignment (detailed in Section \ref{sec:rf-fso-hybrid-los}).

\subsection{Mobile Receiver Units}
The receiver unit can be mobile and needs to be equipped with a PD. It is assumed that the PD(s) are conformal to
the surface of the unit with additional apparatus like lenses as appropriate. These mobiles also need the capability of
uploading using legacy RF transmitters. We are assuming that these mobile devices have one PD receiver and one RF transmitter such as WiFi. They receive the download data from the transmitter(s) with which they are in LOS alignment. The design of these units requires
joint work of solid-state device and packaging as well as communication protocols. For example, multi-element conformal PDs can be designed so that they cover the surface of a smartphone or laptop.

\subsection{RF/FSO Hybrid LOS Management}
\label{sec:rf-fso-hybrid-los}
In our hybrid architecture, the multi-element bulb follows
a software-defined approach to keep track of which receiver
is best aligned with which transmitter. For this, establishing an optical link by associating transmitter(s) to a receiver and maintaining this link with mobility of the
receiver across the room is needed. The controller device also has to partition the transmitters so that multiple transmitters can serve a receiver, and cease the optical link once the receiver is off-line. The LEDs located close to each other (that are in the same partition) are assigned to the same receiver and transmit the same data stream. When a user moves slightly, a handover will not be immediately necessary as it will still be receiving the data signal from the neighboring LEDs in the same partition. Thus, this design will require a handover (or a redoing of the receiver-LED association) only when the receiver makes abrupt movements, which is unlikely inside a room. This protocol allows smooth and continuous mobility for the receivers by electrically steering the data transmission in accordance with the position of the receiver. 
We group these functionality into three basic bulb-mobile association mechanisms as detailed below.

\subsubsection{Establishing the Link}
To search for new receivers in the room, the bulb periodically sends SEARCH frames via its transmitter LEDs. Each LED on the bulb has a local ID, \emph{k}, which is included in the SEARCH frames being sent from that LED \emph{k}. These SEARCH frames are like Ethernet's RTS messages, with a key difference that they are augmented with the local ID of the LED they are being sent from. A mobile receiver \emph{X}, entering the room, receives these SEARCH frames. The receiver might receive multiple of the SEARCH frames depending on its position with respect to the bulb. We assume that the receivers have the capability to filter the SEARCH frame with strongest light intensity. A measure of the received signal strength indication (RSSI) can be fed into the controller where the decision is made over which input has the strongest signal.~\cite{2012-burton-smart}

\begin{figure*}[!htbp]
	\centering
	\subfigure[Varying room size: Square floor dimension increases from 4 m to 20 m. LEDs' divergence angle is 20\degree{} and transmit power is 20mW. The total number of transmitters on the bulb is 25. SIR decreases at a much lower rate  in the region of high interference (depicted by the green arrow) compared to the region of low signal strength (depicted by the red arrow).]{\includegraphics[width=84mm]{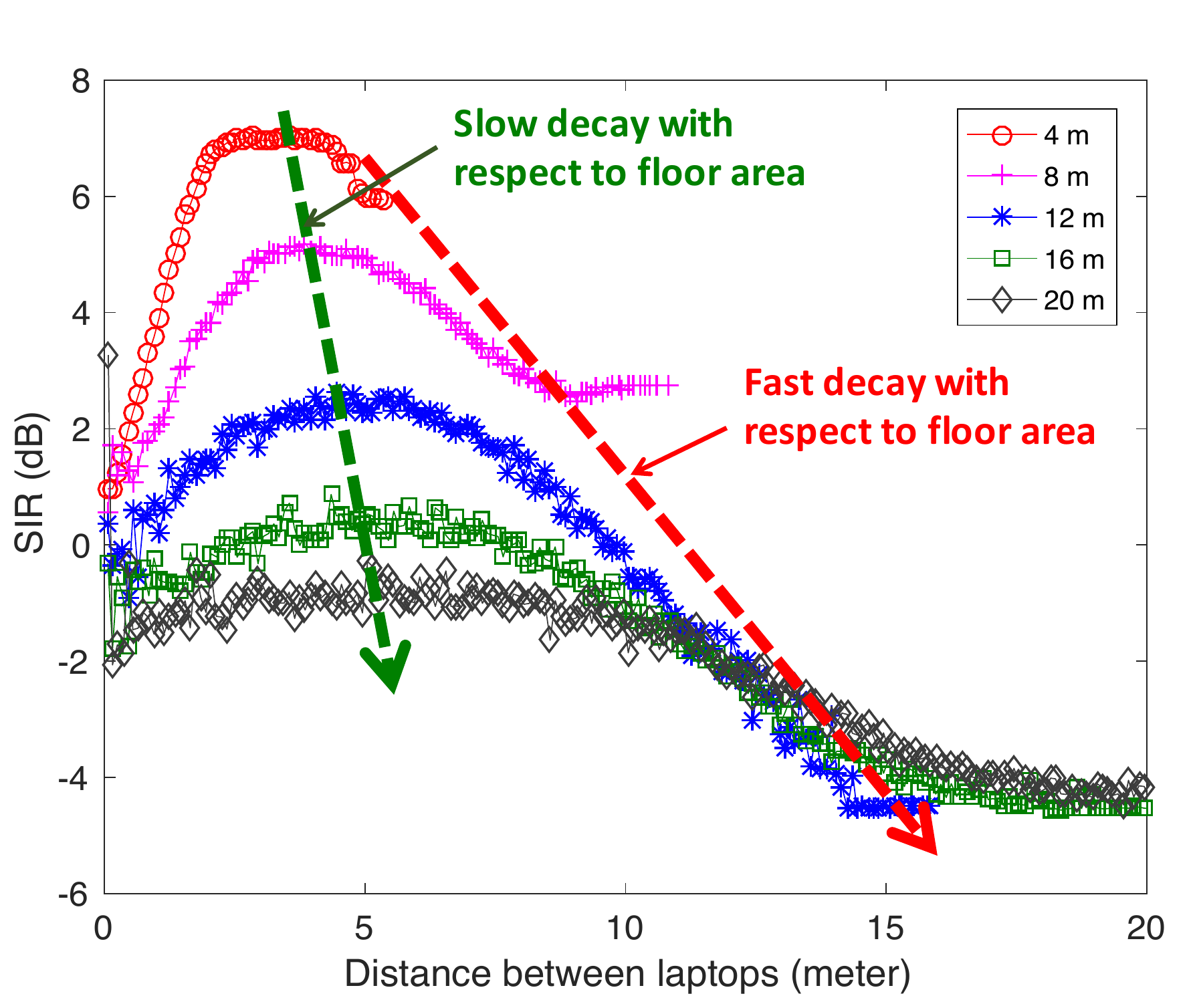}%
		\label{var_room}}
	\hfil
	\subfigure[Varying divergence angle for different amount of total source powers: Square floor dimension is fixed at 6 m. The red ellipse covers the region where the SIR is maximum for different total source powers.]{\includegraphics[width=86mm]{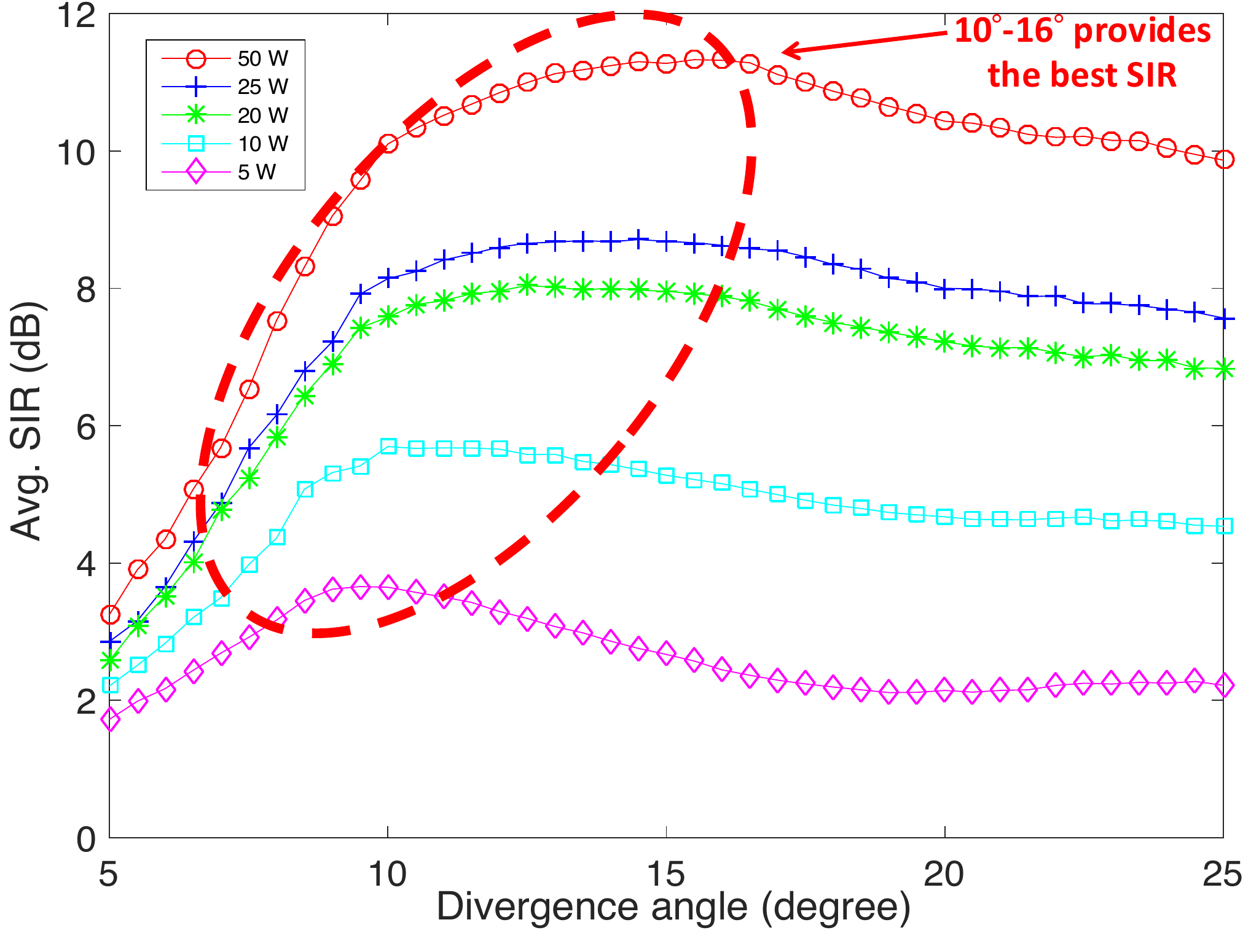}%
		\label{var_dist}}
	\caption{Variation of the SIR with respect to various parameters.}
	\label{SINR_against}
\end{figure*}

Once the receiver receives the SEARCH frame, it sends back an ACK frame (like a CTS in Ethernet) via its RF transmitter. This ACK includes the Ethernet / MAC address of the receiver and the local ID \emph{k} of the LED from which the SEARCH frame was received. The ACK verifies to the bulb that \emph{X} is aligned with the transmitter LED \emph{k}. After receiving the ACK from \emph{X}, the bulb assigns the LED \emph{k} or a group of LEDs around the LED \emph{k} to \emph{X}, and maintains this information as an LED-receiver association table (LED-RAT). When there are multiple receivers in the room, the bulb partitions the LEDs and associates each partition (see Section \ref{sec:design-multielement-bulb}) to a separate receiver. The LED-RAT will need to be updated accordingly. For every data frame to be sent, the bulb does a reverse lookup to the LED-RAT, with the Ethernet address the frame is destined to. In this manner, the bulb steers the data stream destined for receiver \emph{X} onto the transmitters that just got associated to \emph{X}. The partitioning of the LEDs across the receivers will be crucial in the overall performance of the spatial reuse.

\subsubsection{Maintaining the Link}
Once an optical link is established between the receiver
and the LEDs on the bulb, it is maintained by periodic
exchange of SEARCH-ACK messages as described in the previous subsection. When there is a change in the LED-receiver association,
the bulb will need to update LED-RAT and re-partition the
LEDs. Since such changes can happen frequently, it is crucial to keep the complexity of the LED-RAT update and
partitioning of LEDs small. Furthermore, the re-partitioning operation should be performed in a manner independent of the number
of LEDs, as there will be hundreds of LEDs on the bulb.

\subsubsection{Terminating the Link}
When a receiver leaves the room or powers down, the controller in the bulb
needs to update LED-RAT and re-partition the LEDs. There are two possibilities for achieving this:
\begin{itemize} 
	\item
	\noindent \emph{Graceful Leave}: 
	The receiver \emph{Y} lets
	the bulb know that it is powering down by sending a CLOSE
	frame via its RF transmitter. 
	
	\item
	\noindent \emph{Ungraceful Leave}:
	The receiver \emph{Y} simply leaves the room without informing the bulb
	about its departure. Then, the bulb will keep sending
	its SEARCH frames, and will timeout on \emph{Y} after $N_{\rm t}$ SEARCH
	frames without an ACK from \emph{Y}. $N_{\rm t}$ actually indicates the number of search frames without acknowledgement from a receiver after which the bulb will consider the connection between that receiver and itself is timed out. Therefore, $N_{\rm t}$ can be changed under various circumstances.
\end{itemize} 

\section{Design of the Multi-element Bulb}
\label{sec:design-multielement-bulb}

The transmitters of our multi-element hemispherical bulb are arranged in layers of circles to maximize the coverage in the room.
We consider a bulb with multiple layers 
as shown in Figure~\ref{fig_tx_structure}. Efficient arrangement of the transmitter LEDs is within itself an optimization problem that is not discussed in this paper (see e.g.,\cite{murat-Winet} for further discussions on a special
case of this problem). Several factors such as radius and
divergence angles of the LEDs, and height of the room can
affect the light distribution and communication pattern in a
room. An optimized placement should jointly improve the
light distribution and communication in the room.

A heuristic algorithm is devised to partition the LEDs into groups, each corresponding to a mobile receiver in the room,
which takes full advantage of the multi-element bulb for higher
spatial reuse. This reduces the load on LOS alignment
algorithm by providing a wider FOV for each receiver. Further, all LEDs in a partition are modulated with the same transmission signal, and hence, the receiver for that partition can enjoy an aggregate reception quality from the LEDs of its partition. For two receivers positioned at ($X_{1}, Y_{1}$) and ($X_{2}, Y_{2}$) on the room floor, we find the mid
point, ($X_{\rm mid}, Y_{\rm mid}$) and draw an imaginary partitioning line perpendicular to the line connecting the two receiver positions (Figure \ref{fig_part_alg}). Once the partitioning line is settled, then we split the LEDs on the bulb into two categories based on which side of
the line their projections fall. 
This procedure has to be executed every time when a new mobile device establishes connection with a transmitter on
the bulb. In that case, the algorithm will reassign all the
transmitters giving new space for the new connection while
maintaining the old connections. 

\section{Simulation-Based Evaluation}
\label{sec:simulation-evaluation}
In order to get a glimpse of what is possible in terms of
download and illumination efficiency, we first performed an evaluation of our VLC architecture with the partitioning algorithm and coverage model above 
for the hemispherical bulb. Then, we look at the effect of noise on the multi-element VLC architecture.

\subsection{Hemispherical Bulb}
For simulation experiments, 
the multi-element bulb is placed at the center of the room ceiling which is considered as the origin point of the hemisphere. The room height was fixed to 3 meters, and a square floor area is assumed. The hemispherical bulb consists of three layers of transmitters. Each layer is distinguished by the elevation angle between the normal of the hemisphere and the transmitter. In our model, layers 1, 2 and 3 were placed at an elevation angle of 30\degree{}, 45\degree{} and 70\degree{}, respectively. An azimuth angle of 45\degree{} is considered between transmitters. The bulb radius is 40 cm and the transmitter radius is 3.75 cm. We used the simulation parameters in~\cite{sifat-vlcs-2016}, including the light propagation and attenuation models.

\begin{figure}[tb]
	\centering
	\subfigure[Contour plot of SIR versus the distance of both receivers from the center of the floor illustrating the three-region behavior for a bulb with two layers of transmitters. The number of transmitters in the 1st layer and 2nd layer are 11, 17, respectively, while divergence angle of the transmitters is 45\degree{}.]{
		\includegraphics[width=75mm] {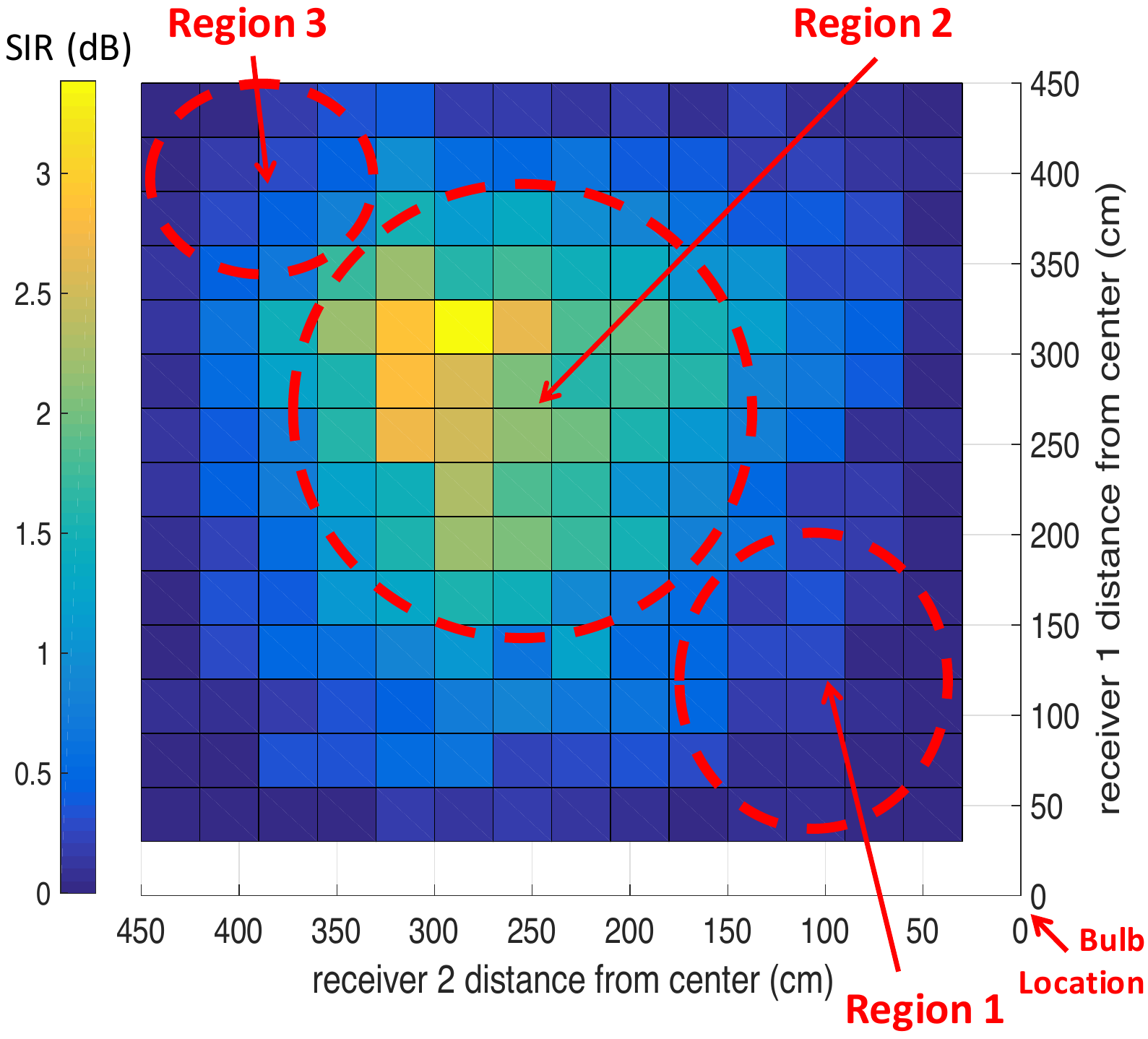}
		\label{fig:regions-1}
	}\\
	\hspace{-2mm}\subfigure[Top view of the room floor illustrating the 3-region behavior in the SIR distribution across the room surface.]{
		\includegraphics[width=70mm] {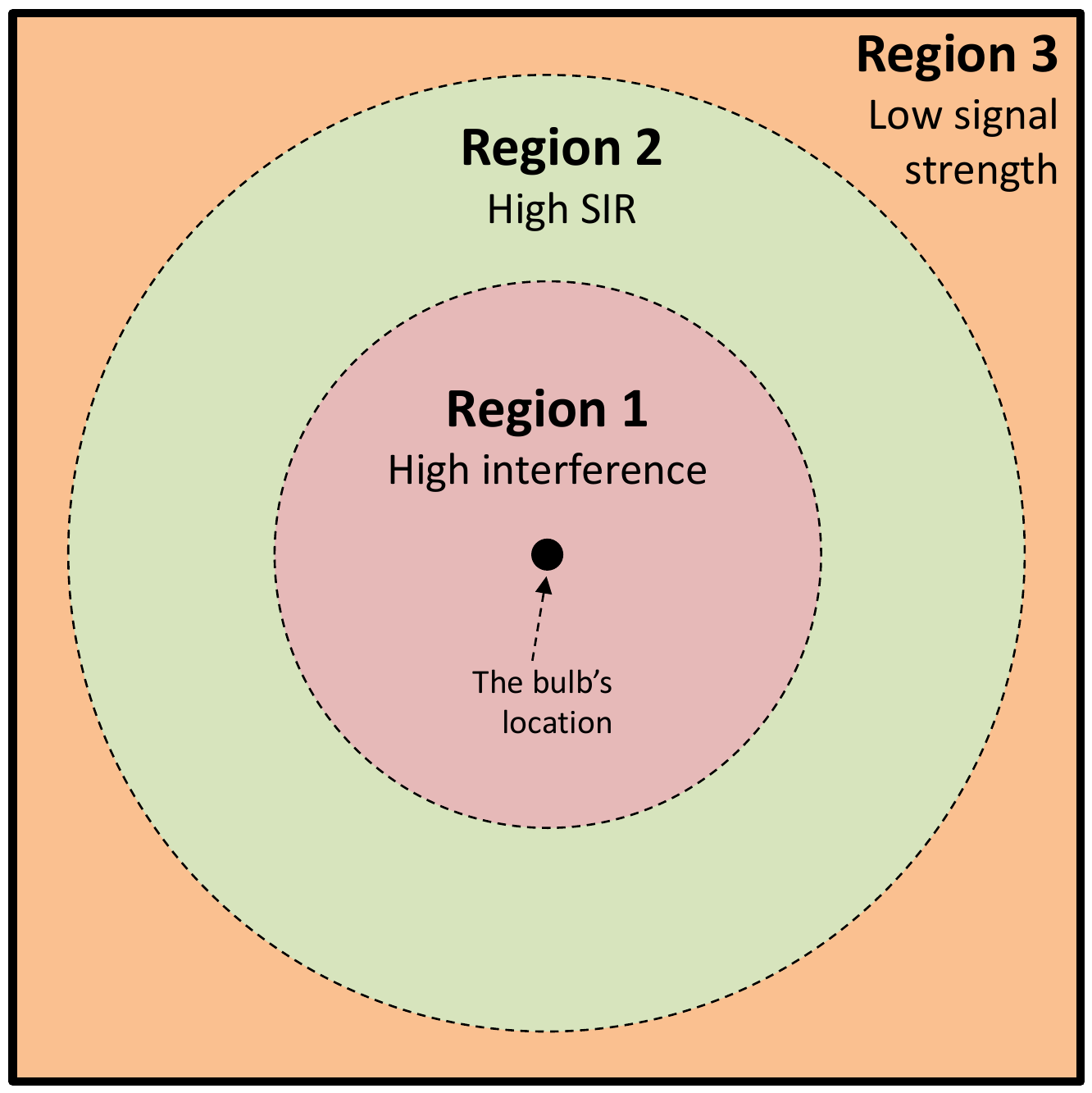}
		\label{fig:regions-2}
	}
	\caption{Three-region behavior.}
\end{figure}

In our experiments, we aim at the optimum arrangement that provides a good signal-to-interference ratio (SIR)\footnote{MATLAB code of the simulations can be found at \textbf{https://goo.gl/O88lkI}.}
To calculate the SIR, we randomly position two receivers on the room floor and divide the LEDs on the bulb into two partitions $P_1$ and  $P_2$ following the heuristic described in Section \ref{sec:design-multielement-bulb}.
We assume that the PD at a receiver is conformal to the
surface of the receiver and can receive the light coming from
a partition on the bulb. The PD is facing upward
to the transmitter and has a radius of 3.75 cm. Light distribution of each LED is simulated based on Lambertian pattern.
Therefore, a receiver is considered to be within the coverage of an LED if it lies
within the divergence angle of the LED. A receiver can be
in the coverage of one LED or multiple LEDs of the same
partition or multiple partitions. Thus, the signals arriving
to a receiver $i$ from partitions other than \emph{$P_i$} need to be
considered as noise. 

For the simpler case of two receivers, we let
$S_{1,1}$ be the total signal received at receiver 1 from LEDs of
$P_1$ and $S_{1,2}$ be the total signal received at receiver 1 from LEDs of $P_2$. Then, SIR for receiver 1 is $\gamma_1$ = $S_{1,1}$/$S_{1,2}$. Following the same notation, SIR for receiver 2 is $\gamma_2$ = $S_{2,2}$/$S_{2,1}$. The interference considered in the SIR calculation is from the light coming from the neighboring transmitters in the
bulb. We do not consider any external noise factors which
should not be very significant in case of a small room. We experimented at three different divergence angles. In the
first two cases, we check the scenarios with small and 
very large divergence angles to test the correctness of the
model. In the third case, we experiment with other parameters to optimize the lighting and signal strength.

\begin{figure}
	\centering
	\hspace{4.2mm}\subfigure[SIR versus Power constraints with and without illumination constraint.]{\includegraphics[width=72mm]{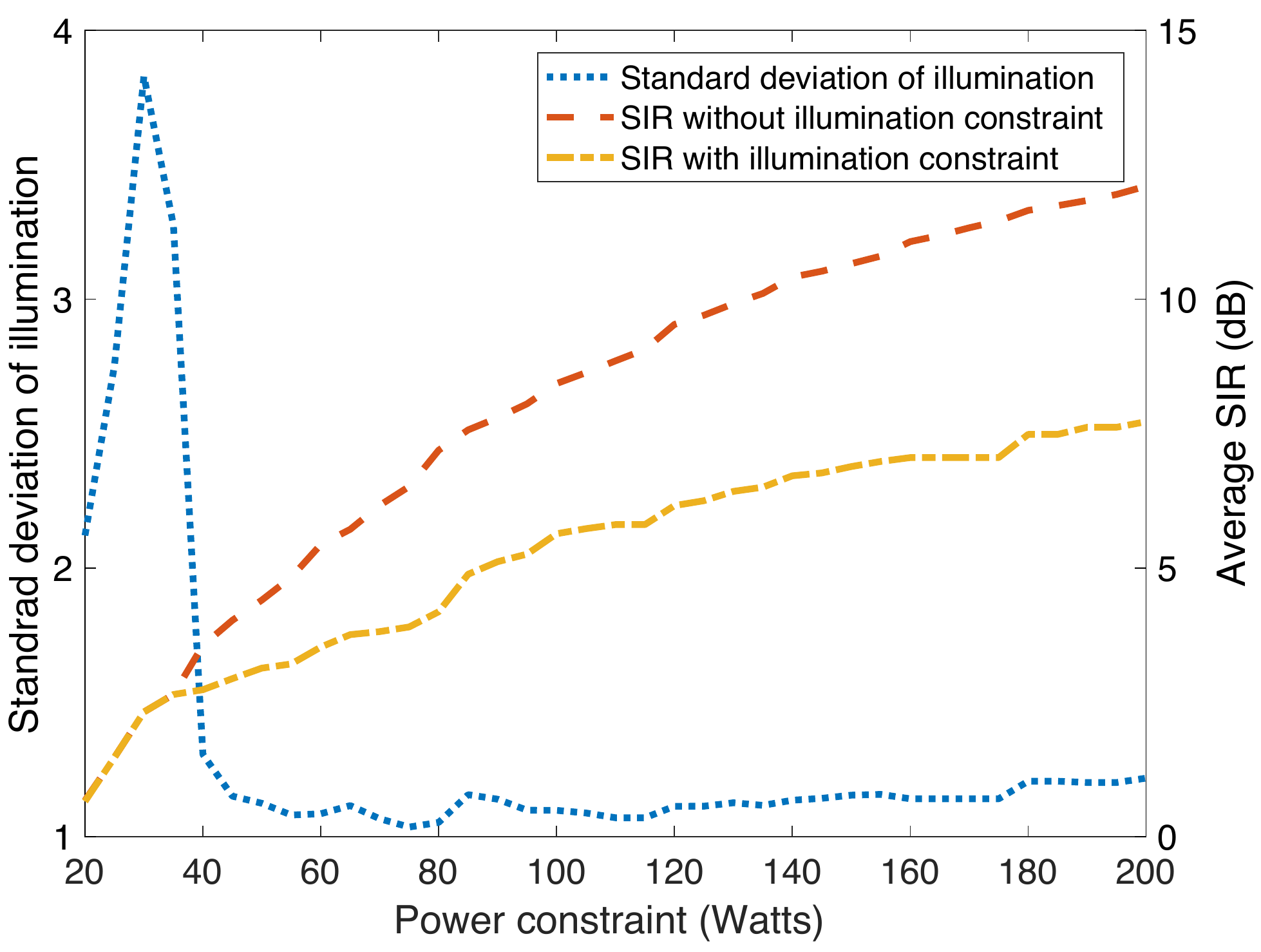}
		\label{fig:SIR-vs-power}}
	\subfigure[Objective function output versus Power constraint.]{\includegraphics[width=69mm]{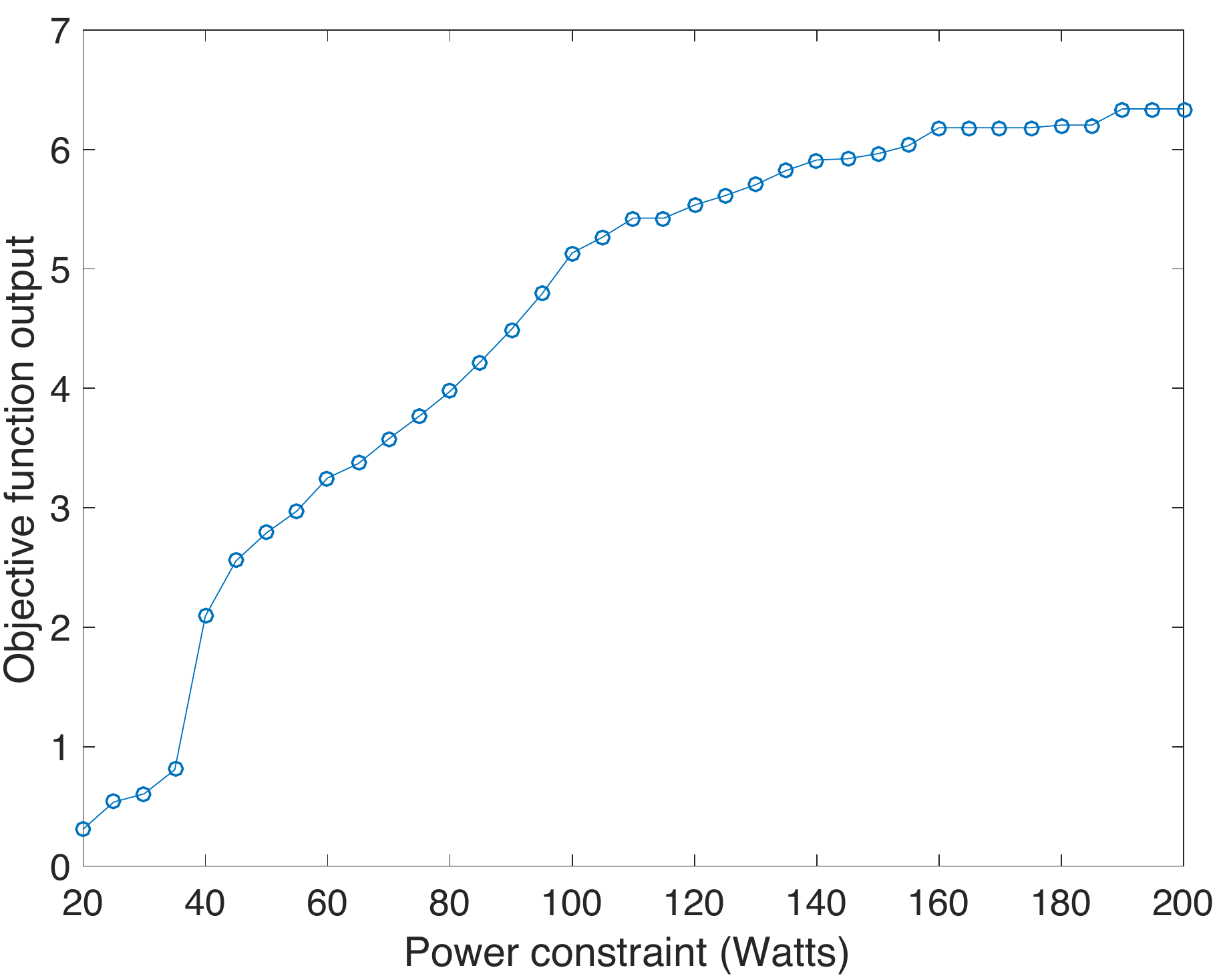}
		\label{fig:SIR-vs-obj}}
	\caption{Comparison of SIR with and without the illumination constraint: The joint optimization finds a balance between the SIR and lighting quality.}
	\label{fig:SIR-vs-power-vs-obj}
\end{figure}

\begin{figure*}[t]
	\centering
	\subfigure[Illustration of 3-LED (top left) and 7-LED (top right) transmitter structures with their layouts in the room (bottom left and right, respectively).]{\includegraphics[width=85mm]{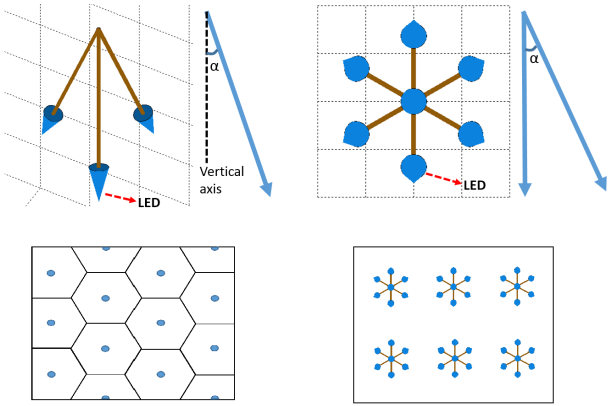}%
		\label{structures}}~~
	\subfigure[CDF of SINR values. In Scenario-1 (S1) we consider that all the LEDs on same transmitter serve the same user, while in Scenario-2 (S2), we consider that all the LEDs on a transmitter serve different users.]{\includegraphics[width=94mm]{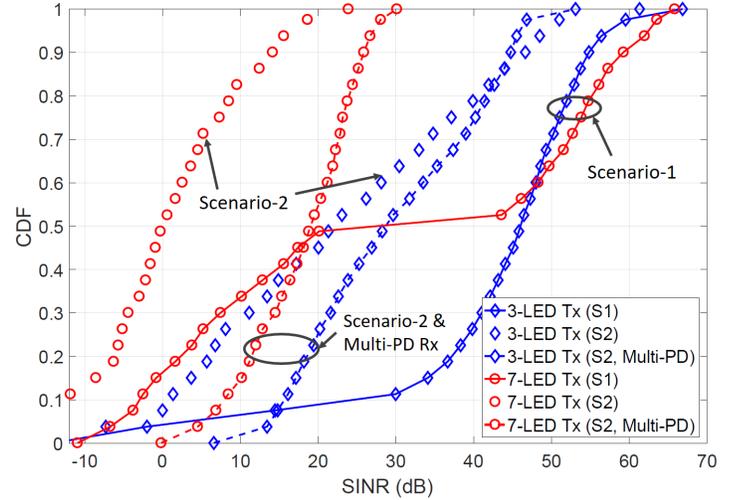}%
		\label{sinr}}
	\caption{Transmitter types and SINR distributions.}
	\label{SINR_CDF}
\end{figure*}

\subsubsection{Varying Room Size and Divergence Angle}
First, to observe the effect of room size, we experimented with varying room sizes using fixed source power for LEDs. The transmit power of LEDs was kept constant at  20mW while the square floor dimension was increased from 4 to 20 meters. As expected, Figure \ref{var_room} shows that the SIR is best when the receivers are neither too close nor too far from each other. 
The SIR deteriorates as the room size (or floor area) increases. This is because for larger room sizes, the likelihood of the two receiver pairs being located farther away from the bulb (e.g., at the corners of the larger room) becomes higher compared to a smaller room.

We also experimented with different divergence angles as divergence angle has a significant effect on the coverage
area of each transmitter, as shown in Figure~\ref{var_dist}. To find the optimum divergence
angle we varied the divergence angle from 5\degree{}
to 40\degree{} for the total source power values of 5 W, 10 W, 20 W, 25 W and
50 W. The SIR results for each total power value were averaged
per divergence angle. As observed from Figure~\ref{var_dist} the optimum divergence angle
is obtained within the range of 10\degree{} and 16\degree{}.

\subsubsection{The Three-Region Behavior}
Figure \ref{fig:regions-1} shows the average SIR of receivers 1 and 2 (i.e., $\frac{1}{2}(\gamma_1+\gamma_2)$) versus distance between each of the two receiver laptops and the floor center, which shows an interesting behavior. The SIRs are averaged over a large number of user locations, each of which satisfy the considered Euclidean distances to the center of the room as in Figure \ref{fig:regions-1}. If the laptops are placed in region 1 (the center region of the room) then the distances between them and the center of the room floor is small, and if we place them in region 3 (in the corners), then this distance has to be large. In the surface plot we can see relatively low values of SIR for small and large distances between the laptops and the center of the room floor and higher values for the medium distances, so this indicates region 2 (the middle region between the center and the corner) to be the most favorable one. In the regions shown in Figure \ref{fig:regions-2}, it is assumed that both of the laptops are inside that region. 
The areas outside the red dotted circles in  Figure \ref{fig:regions-1} indicate cases where one receiver is close to the room center and the other is distant, which also produce low SIR. 

The top view of the surface plot is shown in Figure \ref{fig:regions-1} which more vividly points out the 3 regions (the red dotted circles), since the blue squares indicate low values and the yellow/green squares indicate high values of SIR. The 3 regions in the room floor are shown in Figure \ref{fig:regions-2}. Also in Figure \ref{var_room}, it
can be observed that SIR in the high interference region 1
reduces much slower than the floor area, and in region 3, SIR
decreases along with the floor size. This behavior can act as a useful guide in organizing the room layout for the placement of the receivers.

\subsubsection{Optimum Bulb Design Under Illumination Constraint}
\label{sec:optim}
We varied the number of transmitters in each layer and the transmitters' divergence angle to find the optimum bulb design considering both illumination and signal strength. 
We first explored optimum bulb configurations for a particular number of layers to see which combination produces the maximum SIR under an aggregate power constraint. We solved this optimization problem (more details about problem statement are in~\cite{sifat-vlcs-2016}) by varying the number of LED boards per layer and the divergence angles of LEDs. We then updated this problem by adding an illumination constraint, which is the standard deviation of illumination across the room floor, and the objective function of this optimization problem is defined as the SIR divided by this illumination variance. We have compared the results of these two scenarios under various power constraints.

Figure \ref{fig:SIR-vs-power} shows optimum SIR versus power constraint, which shows a rise at the beginning, but as expected, the best achievable SIR saturates as the power constraint increases. It is also observed that the optimum SIR is significantly lower when there is an illumination constraint, again confirming our expectations. For the illumination variance, it gradually decreases as the power constraint increases, since better configurations are possible at higher power constraints. In Figure \ref{fig:SIR-vs-obj} the plotted objective function also gets saturated, because when the power constraint is increased (that is, the number of LED boards in a configuration can be increased) it means that the signal strength is also increased. Since we cannot have unlimited number of transmitters on the bulb, after a certain value of the power constraint, we cannot get any more improvement in the SIR. Even with maximum possible number of LED boards in each layer, we cannot achieve the best SIR because of higher interference coming into action. Therefore, after that threshold value the maximum possible SIR value remains the same. 

Since the main objective is to find a good balance between the SIR and the lighting quality, this is clearly evident from Figure \ref{fig:SIR-vs-power-vs-obj}. We can see that from every value of the power constraint, SIR considering the illumination constraint is lower than the SIR without considering the constraint because of the low value of the variance (which indicates better quality of lighting), and that is why the value of objective function is the highest. 

\subsection{SINR Distribution Across the Room}

To understand the effect of different transmitter structures on the VLC link quality, we evaluated signal-to-interference-plus-noise-ratio (SINR) performances of two different multi-element transmitter structures; 3-LED and 7-LED transmitters. Transmitter structures of multi-LED transmitters and their layout in the room are shown in Figure~\ref{structures}. A large room of size 15~$\times$~17~$\times$~4~m$^3$ is considered for simulations, and transmitters are located at ceiling height. Up to four wall reflections are considered to generate multipath realizations. For a fair comparison between different structures, we deployed 14 of 3-LED transmitters for the first configuration, and 6 of 7-LED transmitters for the second configuration as shown in Figure~\ref{structures}. The half-intensity radiation angle of the LEDs is 30\degree and the PD's FOV is 40\degree. We captured the SINR at large number of points in the room, by slightly changing location of the receiver each time, and presented the results as cumulative distribution function (CDF). 

We considered two particular scenarios for capturing the SINR distribution within the room which are explained in detail in~\cite{7414124}. In the first scenario (scenario-1), we assumed all the LEDs on the same transmitter serve the same user. In this case, all the signals coming from a transmitter constructively add up, and yield a good SINR. Drawback of this scenario is, although the transmitter has many elements it only transmits a single data stream. SINR CDFs of scenario-1 is shown with solid line and markers in Figure~\ref{sinr}. Normally, we expect 7-LED transmitter configuration to give a better SINR distribution, because more LEDs serve a single user. However, in the illustration we see that SINR is not proportionally increased, especially at low SINR region. The SINR CDF of 7-LED transmitter is split into two regions, a high SINR region and a low SINR region; and there is a large gap between those regions. This condition is mostly caused due to non-hexagonal layout of 7-LED transmitters in the room, which results in the users at cell edge to observe high interference and yield low SINR. At high SINR region, 7-LED configuration have slightly better SINR values compared to the 3-LED configuration. These high SINRs are reached within the area immediately below the transmitters, where the received signal strength is strong and interference power is small. The CDF of 3-element configuration is smoother than 7-LED configuration, which implies a more uniform coverage distribution in the room. 

In the second scenario (scenario-2), we assumed that all the LEDs on a transmitter will serve different users, which means they will all transmit different data streams. In this case, LEDs will cause interference to each other, and in total this system will yield a lower SINR for any user. Advantage of this scenario is, since each transmitter will serve many users simultaneously, aggregate throughput will be much higher than scenario-1, which is left as a future study. The SINR distribution for the scenario-2 is shown with no line, only markers in Figure~\ref{sinr}. As all the other LEDs at a given transmitter are assumed to be interference sources (since they are serving to different users), this scenario yields low SINR geometries. When the number of LEDs on a transmitter increases, SINR decreases dramatically. The reason is, since LEDs are closer to each other, they cause significant cross-interference. Second scenario serves to more users at the same time, but needs higher SINR for high speed data transmission.

To deal with low SINR problem of scenario-2 and narrow the SINR gap between scenario-2 and scenario-1, we considered using diversity combining techniques at the receiver with 7~PDs and optimal combining. The CDF results with multi-element receiver is shown with dashed line along with markers. With the use of receiver diversity and optimal combining, a 20~dB gain is observed at 7-LED transmitter configuration and 15 dB gain is observed at 3-LED configuration for the median SINRs. With the receiver diversity improvement, multi-element transmitter can both serve multiple users and provide good SINR.

\section{Concluding Remarks}
\label{sec:summary}

We introduced a multi-element VLC architecture that employs a hemispherical bulb with multiple narrow FOV LEDs. The mobile receivers use VLC for download and RF for upload, and the multi-element bulb uses a software-defined approach to manage LOS alignment with receivers. We modeled the bulb structure, and presented preliminary results showing that the architecture can offer high spatial reuse while keeping a desired illumination level.
Also, we presented a framework for optimizing the multi-element bulb design not only taking the signal quality into consideration but also the evenness of lighting across the room.
We believe that the software-defined VLC framework will greatly contribute to the field of VLC, particularly for the IoT applications. 

The presented framework can serve as a basis for future studies for better understanding and further improvement. For instance, studying multiple bulbs in a room would lead improvement in the illumination and communication quality. 
Moreover, investigation of other shapes (triangular, square etc.) in addition to the studied hemispherical shape could provide insights on ideal bulb geometry. The algorithm for partitioning LEDs among receivers can also be further improved to better balance time/computational complexity and higher spatial reuse opportunities.

\ifCLASSOPTIONcaptionsoff
  \newpage
\fi

\bibliographystyle{IEEEtran}
\bibliography{./new}  

\end{document}